\documentclass[twocolumn]{aastex702}

\usepackage{newtxtext,newtxmath}
\usepackage{multirow}

\usepackage{graphicx}	
\usepackage{amsmath}	
\usepackage{subcaption}

\begin{document}

\title{An Inclined, Eccentric Planet and an Inner Debris Disk Could Reproduce AU Mic Structure}


\author{Arcelia {Hermosillo Ruiz}}
\affiliation{Physics and Astronomy, University of Exeter}
\affiliation{Astronomy and Astrophysics, University of California Santa Cruz}
\email[show]{a.hermosillo-ruiz@exeter.ac.uk}

\author{Ruth Murray-Clay}
\affiliation{Astronomy and Astrophysics, University of California Santa Cruz}
\email{rmc@ucsc.edu}
\author{Meredith A. MacGregor}
\affiliation{Physics and Astronomy, John Hopkins University}
\email{mmacgregor@jhu.edu}
\author{Renata Frelikh}
\affiliation{Laboratory for Atmospheric and Space Physics, University of Colorado Boulder}
\email{Renata.Frelikh@lasp.colorado.edu}

\begin{abstract}
The debris disk orbiting the M star AU Microscopii has a series of large-scale clumps that move away from the star at high velocities above the mid-plane on the southeast side. Two more bright features lie on the northwest side of the disk, localized below the mid-plane and moving toward the star. These clumps are only observed in scattered light indicating that they affect small 0.2$\mu$m-sized grains. We present a mechanism for emitting periodic dust clumps by appealing to stellar forces and an inclined, eccentric planet interacting with an exterior debris disk. In our best-matching simulations, the planet exerts an impulse on the disk every orbital period, generating periodic enhancements in dust above the mid-plane. We assume that the stellar wind only acts on grains once they reach a height above the mid-planet that exceeds a threshold value (a free parameter in our model), at which point they are accelerated outward. The behavior of periodic particle ejections and trajectories depends significantly on the planet's mass, eccentricity, and inclination; separation between the planet and disk; and the ratio of stellar wind force to the star's gravitational force ($\beta$). We find a promising qualitative match to observations with simulations that include an as-yet-undiscovered and observationally allowed planet with mass $2 M_J$, semi-major axis between 3-4 au, eccentricity of 0.37, and inclination of 30$^\circ$, a ring of particles between 5-6 au, and a stellar wind height threshold of $z_h = hr$, where $h \approx 0.02$. We visualize our simulation with surface brightness maps to compare with existing observations of AU Mic. We find that a value of $\beta \approx 1.8$ accelerates the clumps radially outward at velocities that are comparable to the clumps seen in the AU Mic disk and produces features similar to those observed. 
\end{abstract}

\keywords{keyword1 -- keyword2}



\section{Introduction}

In the late stages of planet formation, once most primordial gas has dissipated, a circumstellar ``debris disk'' composed of dust and planetesimals remains 
\citep[see reviews by][]{Wyatt2008,Matthews2014,Hughes2018}. 
Collisions between kilometer-sized or larger planetesimals initiate a collisional cascade that forms a second generation of dust and continuously replenishes the disk \citep{Strubbe2006}. Approximately mm-sized particles in the collisional cascade allow debris disks to be observable at (sub)-millimeter wavelengths, where the disk's thermal emission is significantly larger than that of the stellar photosphere. In the optical, scattering of starlight by $\sim$$\mu$m-sized grains dominates. 

The edge-on debris disk around the 23 $\pm$ 3 Myr \citep{mamajek2014}, M-type star, AU Microscopii (often referred to as AU Mic in the literature and for the remainder of this paper), was first imaged by \citet{Kalas2004} at optical wavelengths. A peak in vertical optical depth indicates that most of the disk's mass is located in a thin ring at $\sim$ 35 au \citep{Augereau2006}.  While \citet{Strubbe2006} suggested that the region interior to this thin planetesimal belt may be devoid of micron-sized grains, with observed emission at these radii arising from forward scattering by grains in the belt, ALMA observations suggest the presence of a collisional disk with a substantially wider extent \citep{MacGregor2013,Daley2019}.  The inner radius of the disk seen at mm wavelengths was constrained to a 3$\sigma$ lower limit of 5 au and upper limit of 20 au \citep{MacGregor2013,Daley2019}.  \citet{Vizgan2022} identified an additional interesting disk feature---they measured an aspect ratio at $\lambda=$ 450 $\mu$m that is smaller than that at $\lambda=$ 1.3mm \citep{Daley2019}, indicating that smaller dust grains are less extended in the vertical direction. This result suggests that non-gravitational forces are important for the dynamics of these particles because the smaller dust particles are more susceptible to stellar wind forces, thus leading to a larger expected disk scale height \citep{pan2012,Quillen2006}. 

\begin{figure*}
    \centering
    \includegraphics[width=6in]{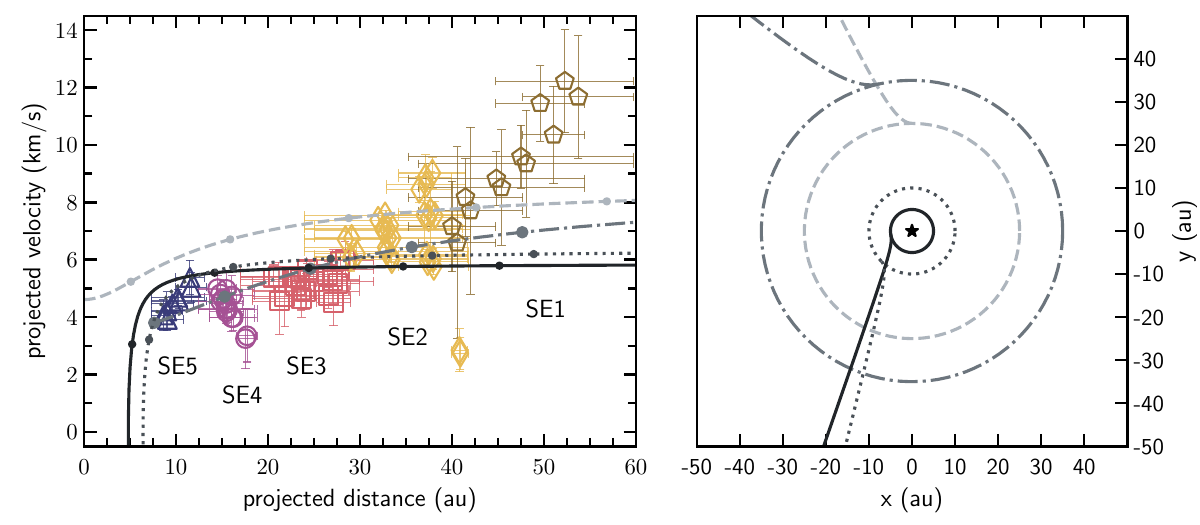}
    \caption{We simulate four particles with a different combination of initial distance from the star ($R_d$), stellar wind force ($\beta$), and ejection angle ($\theta$), demonstrating how these parameters affect the projected speed at a given projected distance; line styles are consistent across panels. \textit{Left:} Projected speed and projected distance of the four particles and the five southeastern dust clumps  SE5 (blue triangle), SE4 (purple circle), SE3 (pink square), SE2 (yellow diamond), SE1 (brown pentagon), calculated from Table 2 in \citet{Boccaletti2018}.  Each trajectory passes through the data points of at least two clumps, illustrating that particles on hyperbolic orbits due to stellar wind forces can reproduce the observed projected speeds at a range of distances, though no single trajectory fits all five clumps simultaneously. Filled circles along each trajectory are separated by 8.5 years; note that these circles align with the projected distances of the observed clumps, motivating the idea that a planet with an $\approx 8$-year orbital period drives the periodic ejections. There is no planet included in these simulations.
    \textit{Right:} A top down view of the same four particles in their initial circular orbits and their subsequent hyperbolic trajectories. Ordered from smallest to largest initial orbit, the parameters that determine the trajectories ($R_d$, $\beta$, $\theta$) are: (5 au, 2, 1.06$\pi$), (10 au, 7, 5$\pi$/4), (25 au, 10, $\pi$/2), (35 au, 5, 1.14$\pi$/2), where $\theta=\omega$ since $\Omega = f= 0$. The ejection angle $\theta$ determines the direction in which particles are launched, which affects the projected speed as seen by an observer.}
    \label{fig:velocity_position_data}
\end{figure*}

AU Mic gathered attention due to five bright, fast-moving scattered-light features localized in the southeast (top left\footnote{For ease of interpretation, we refer to the southeast direction as ``left" and the northwest direction as ``right," corresponding to the orientation of the plots in \citet{Boccaletti2015} and this work.}) side of the disk at projected separations from the star of $\sim$10-55 au \citep{fitzgerald2007,schneider2014,Boccaletti2015}. These clumps of dust have been resolved in scattered light images taken with HST/STIS and VLT/SPHERE over 10+ different epochs between 2004 and 2017 \citep{Boccaletti2015,Boccaletti2018}. The features are intriguing because the majority of the puffs appear on one side of the disk and have velocities between $5-15$ km/s, where the clumps at large projected distances, exceed the system's escape velocity (see left panel in Figure \ref{fig:velocity_position_data}). The observations between 2015-2017 also found two features on the bottom right with a potential new southeast feature that is yet to be confirmed with new observations \citep{Boccaletti2018}. The two features closest to the star are brightest and reach heights of $\sim 1$ au, and the last three features are too diffused to strongly constrain their elevation from the mid-plane, though they appear to decrease in height over time. A few models have been proposed to explain these features \citep{Sezestre2017,Chiang2017}, but none are able to reproduce the properties of the system entirely so the mystery remains.   

Gravitational interactions between planets and disk material affect a disk's radial, vertical, and azimuthal structure \citep[e.g.,][]{pan2012,pearce2024}, and so asymmetries in debris disks have motivated probing the presence of planets \citep{Marino2018,Booth2023}. In some cases, these studies actually led to finding otherwise-undetectable planets \citep{Lagrange2009,Lagrange2025}. So far, no planet has been found that can be linked to creating the clumps in AU Mic. However, AU Mic does host a planetary system. Three close-in planets have been confirmed via transits and transit timing variations, with orbital periods 8.46, 12.73, and 18.86 days \citep{Plavchan2020,Martioli2021,Wittrock2023}. A fourth planet candidate was detected through the radial velocity method and is suggested to have an orbital period of 33.39 days, though follow-up work is necessary to validate this \citep{Donati2023}\footnote{Updated planet parameters can be found in the exoplanet archive: \href{https://exoplanetarchive.ipac.caltech.edu/overview/aumic}{https://exoplanetarchive.ipac.caltech.edu/overview/aumic}}. All planets have semi-major axes $\lesssim$ 0.15 au, which is well within the inner radius of the debris disk as constrained by ALMA. Planets more massive than 3 Jupiter masses at separations $\gtrsim$ 3 au would have $\gtrsim$ 50\% probability of detection with JWST \citep{Lawson2023}, meaning planets of $\lesssim2$ Jupiter mass around a few au remain plausible with current observational upper limits.

Non-gravitational forces also affect the structure of a disk; stellar forces -- notably radiation pressure and stellar winds -- can displace or completely clear out small dust grains \citep{Burns1979,schuppler2015}. Both radiation pressure and stellar winds are proportional to and act against the gravitational force; therefore, these two forces are often combined into an effective force and parameterized by $\beta = \beta_{\rm wind} + \beta_{\rm rad} = \frac{F_{\rm wind} + F_{\rm rad}}{F_{\rm grav}}$, where $F_{\rm wind} $ is the stellar wind force, $F_{\rm rad}$ is the radiation pressure, and $F_{\rm grav}$ is the gravitational force. Particles released from circular orbits will be blown out of the system when $\beta \ge 0.5$. We note that stellar winds dominate clearing of small dust grains in AU Mic because the star is very active, and its luminosity is too low to produce significant radiation pressure force \citep{Arnold2019}. In our work, we treat $\beta \approxeq \beta_{\rm wind}$. Estimates for the properties of AU Mic's stellar wind suggest mass-loss rates between $10\dot{M_\odot} - 300 \dot{M_\odot}$ and even as high as 2500 $\dot{M_\odot}$ during high activity \citep{Augereau2006,Strubbe2006,schuppler2015}. Wind speeds between 400 km/s and 2200 km/s have also been used in the literature \citep{Augereau2006,Strubbe2006,schuppler2015,AlvaradoGomez2022}.

Two models have been proposed to explain the source of dust for the AU Mic puffs and the mechanism for producing them. \citet{Sezestre2017} performed numerical simulations of test particle trajectories to find best-fit values for the particle's radial locations at the time of ejection and $\beta$. They considered two scenarios, one where the source of dust is at a fixed location, and another where the dust source is in a Keplerian orbit. Their fits suggest a dust production source at $\approx 8-28$ au and $\beta$ values of $\approx$ 10.  \citet{Chiang2017} appealed to dust avalanches where dust grains are accelerated by the stellar wind from a location that intersects the dust-rich birth ring (35 au) and a hypothetical inclined secondary ring, which is a remnant from a catastrophic collision involving a $400$ km-sized planetesimal. They performed numerical simulations of dust grains initialized at a radial location of 35 au and experiencing stellar wind forces, where $\beta =$ 20 and 40 produced the best matches to the observations. Their model assumes that either the stellar wind is variable and pushes dust grains at a periodicity that matches the data, or the avalanches are triggered when some threshold condition is periodically satisfied in the intersection region. These two models have noted the periodicity of the clumps and relied on hyperbolic orbits to interpret the observations. 

Building on these previous models, and motivated by the dust features' periodicity, we propose a new mechanism for creating the escaping clumps of dust in the AU Mic system. We draw attention to the periodicity of the clumps (7-10 yrs) and we suggest that a so-far-undiscovered planet with an orbital period between 7-10 yrs may be part of the equation for solving this mystery. In our model, an eccentric, inclined planet perturbs an exterior disk and produces periodic puffs; therefore, the particles would originate from an inner region of the disk ($\gtrsim 5$ au). Due to the presence of mm-size grains as close as 5 au, by default, we expect smaller grains to exist in this region due to collisional processes.

We only consider cases where the planet's orbit does not cross that of the debris disk, but the eccentricity is high enough for the planet to perturb the dust particles when it approaches apocenter. At apocenter, the inclined planet kicks a group of particles closest to the planet, launching the particles above the mid-plane, where stellar radial forces accelerate small dust particles outward. We consider three scenarios at this stage: either self-shielding keeps mid-plane particles from being blown out as fast as particles that land above the scaleheight of the disk, all $\sim$micron-sized grains are affected by the stellar forces equally and the steady flow in the mid-plane is such that the disk does not appear to change much, or small grains are created at each planet passage through collisions between planetesimals that the planet perturbs. This process repeats every time the planet reaches apocenter and periodic puffs of dust are emitted each orbital period. 

In Section \ref{sec:model} we motivate the model and explain how we constrain the parameter space using the impulse approximation and Nbody simulations. In Section \ref{sec:large_scale_simulation}, we visualize the disk with surface brightness maps and show that we produce promising qualitative agreement with observations. We discuss our assumptions and implications for future work in Section \ref{sec:discussion} and summarize our results in Section \ref{sec:summary}.

\section{Model}
\label{sec:model}

\subsection{Qualitative Motivation}

We interpret these bright features in a similar way as \citet{Sezestre2017} and \citet{Chiang2017}: 1) the features are ``clumps'' of dust accelerated radially away from the star due to stellar forces and 2) the clumps are ejected periodically. Furthermore, we allow ejected particles to move toward or away from the observer's line of sight, though we find best results when the particles making the clumps follow the same path (see Section \ref{sec:large_scale_simulation}). If we assume the five southeast dust clumps were emitted periodically at equal length intervals, from the same radial location, via the same mechanism, and with a constant outward radial force, then we expect each new dust clump to follow the same trajectory as the older ones. In this scenario, the oldest dust clumps are those farthest from the star and the newest ones are those closest to the star. For reference, the projected speeds of the 5 AU Mic features (SE1, SE2, SE3, SE4, SE5; ordered from farthest to closest to the star and using the nomenclature from \citet{Boccaletti2018}) are shown in Figure \ref{fig:velocity_position_data} which are derived from the data in their Table 2.

Given our interpretation described above, particles with distinct hyperbolic orbits are capable of fitting the AU Mic clumps' velocity data. In other words, particles can be ejected from different distances, travel at different radial velocities due to varying magnitudes of stellar forces, and have similar projected speeds for a given projected distance. The three parameters that affect particle trajectories are particle distance from the star, angle when particles are blown away, and $\beta$. We show how four different combinations of parameters produce distinct particle trajectories that similarly match the data in Figure \ref{fig:velocity_position_data}.  
For this example, and every simulation shown in this paper, we integrate the trajectories of the particles with the Nbody code, \texttt{REBOUND} \citep{Rein2012,Rein2019} and use the module \texttt{radiation\_forces} within \texttt{REBOUNDx} \citep{Tamayo2020} to assign a $\beta$ value for each particle.
We initialize the 4 massless particles, orbiting a $0.6 M_\odot$\footnote{Note, this is the mass of the AU Mic star we adopt in this work.} star, with semi-major axes ($a$) = 5, 10, 25, 35 au, arguments of pericenter ($\omega$) = 1.06$\pi$, 5$\pi$/4, $\pi$/2, 1.14$\pi$/2, and all other orbital elements -- eccentricities ($e$), inclinations ($i$), longitudes of ascending node ($\Omega$), and true anomalies ($f$) -- are equal to 0. Each particle is assigned $\beta$ = 2, 7, 10, 5, respectively (right panel in Figure \ref{fig:velocity_position_data}). In these examples, $\omega$ is analogous to the angle at which particles are ejected, $\theta$, which is the angle relative from a line at $y = 0$. 
The particles' hyperbolic trajectories depend on $\theta$, so the initial values of $\omega$ are very specific. The particle orbits on the right panel  of Figure \ref{fig:velocity_position_data} have the same line style with their respective particle on the left panel.

Motivated by these particle trajectories and the periodic nature of the clumps, we consider that a hidden planet with the same period is responsible for emitting a coherent group of particles each orbit. We represent the orbital period of a planet with the filled circles along the particle trajectory curves, where each period lands near each clump (left panel Figure \ref{fig:velocity_position_data}). The period represented here is 8.5 years. 

\subsection{Planet-Disk Model}
\label{sec:planet_disk_model}
\begin{figure}
    \centering
    \includegraphics[width=3.3in]{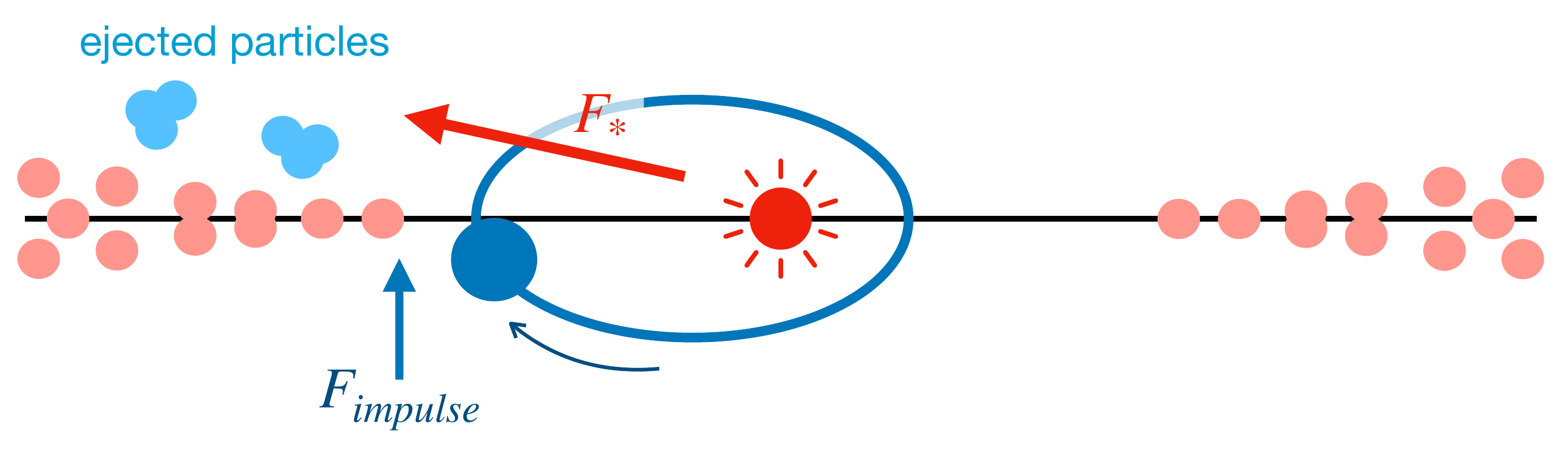}
    \caption{A not-to-scale schematic of an eccentric, inclined planet and exterior debris disk model proposed in this paper. The planet imparts a larger vertical kick on particles when it reaches apocenter, compared to elsewhere in the disk. This kick occurs once every orbit and a new dust clump is kicked up every orbital period; therefore, the periodicity of the clumps match the planet's orbital period. The clumps move outward due to the stellar wind force. }
    \label{fig:modelcartoon}
\end{figure}

We propose a model with an inclined, eccentric planet interior to a dust-rich debris disk. The eccentricity should be high enough so that when the planet reaches its apocenter, it perturbs the dust particles with a larger force than anywhere else in the disk. We only consider planetary non-crossing orbits. Because the planet is inclined, it will impart a vertical velocity onto dust particles when it approaches apocenter. The force the particles feel, and therefore the height the particles reach, due to the impulse from the planet, depends on the planet's mass, eccentricity, and inclination and radial separation between the planet and disk. Through this gravitational interaction between the planet and disk, a dust clump is produced above the mid-plane. Then, the stellar wind is responsible for radially accelerating the clump away from the star. This process repeats and new clumps are generated once every orbital period. This concept is shown in a not-to-scale schematic in Figure \ref{fig:modelcartoon}. 

We consider three possibilities to explain how dust grains above the mid-plane are pushed outward without completely clearing out the material in the mid-plane. In the first scenario, we expect particles that reach heights above the scaleheight of the disk are more susceptible to being blown out by the stellar wind. In the mid-plane, there is a higher density of material and the grains and bodies shield each other from the stellar forces so they are not pushed out entirely. Above the mid-plane, this self shielding no longer occurs so the $\mu$m-sized grains are affected by the stellar radial forces more easily. An alternative possibility is that all sub$\mu$m-sized particles, regardless of their vertical distance from the mid-plane, feel the same magnitude of stellar wind force. Dust in the mid-plane accelerates constantly outward, but we do not see it moving because the flat structure does not change over time. A third option is that at every planet's close approach to the disk, large grains collide to produce the sub$\mu$m-sized grains that are then ejected as a clump. In Section \ref{sec:height_beta}, we discuss the validity of these three scenarios based on order of magnitude calculations, and we find that the last two scenarios are favored but the first also merits continued consideration.

For this paper, we are agnostic regarding which scenario may be responsible for enabling radially accelerating sub$\mu m$-sized grains above the mid-plane while retaining material in the mid-plane. In our simulations, we choose to turn on stellar forces for particles as soon as they reach a certain elevation, which we leave as a free parameter in the model and is denoted by $z_{\rm h}$. We implement this feature by tracking the height of the particles every timestep ($z_{\rm pt}$) and assigning a value for $\beta$ (i.e., essentially ``turning on'' the stellar wind) if $|z_{\rm pt}| > z_{\rm h}$ or assigning $\beta = 0$ (turn off stellar wind) if $|z_{\rm pt}| < z_{\rm h}$. By following this method, we do not require particles to be replenished in our simulations.

\subsubsection{Dynamics of the planet-disk model }
\label{sec:test_model}

We explore the dynamical behavior between an inclined, eccentric planet and an exterior ring of particles to constrain the parameter space in which a planet kicks particles to produce periodic ejections from the disk. The parameters that affect the dynamics of this system and particle trajectories are: the planet's semi-major axis ($a_P$), eccentricity ($e_P$), inclination ($i_P$), mass ($M_P$); the particles' radial location ($R_d$, disk inner edge); the height when particles feel stellar wind ($z_h$); and the stellar wind force ($\beta$).

We explore parameter space with Nbody simulations of a planet, and 200 massless particles. 
The goal of these pilot simulations is to understand how particles will behave under the gravitational influence of an eccentric, inclined planet and how their trajectories will look due to distinct initial conditions. Here we discuss the parameter space we explore in our test simulations and the values we arrive at.

We start with arbitrary values for the planet and disk inner edge to check that the planet kicks the particles to a height comparable to that assumed by an impulse approximation. 
Assuming $M_P<<M_*$, the velocity of a planet in an elliptical orbit is $v_P^2 = GM_*(2/r_P-1/a_P)$, where $G$ is the gravitational constant, $M_P$ is the mass of the planet, $M_*$ is the mass of the star, $r_P$ is the radial distance of the planet, and $a_P$ is the semi-major axis of the planet. The apocenter distance is $a_P(1+e_P)$. The velocity at apocenter is then $v_{P,\mathrm{apo}} = \sqrt{\frac{GM_*}{a_P}\frac{(1-e_P)}{(1+e_P)}}$.  
The impact parameter for an encounter between the planet and a disk particle is $b = a_l - a_P (1+e_P)$, where $a_l$ is the semi-major axis of the disk particle before an encounter. Particles initialized with zero inclination are kicked by a planet. After the encounter, the particles reach a height given by $z_\mathrm{pt} = a_{l}\frac{v_\mathrm{rand}}{v_K}$, where $v_\mathrm{rand}$ is the velocity of the disk particle in the vertical direction that is kicked up by the planet and $v_K$ is the Keplerian velocity. From the impulse approximation,
$v_\mathrm{rand} = \frac{GM_P}{b^2}\Delta t =  \frac{GM_P}{b v_\mathrm{rel}}$, where $\Delta t = b/v_\mathrm{rel}$, and $v_\mathrm{rel}$ is the planet's relative velocity in the vertical direction given by $v_{P,\mathrm{apo}} sin (i_P)$. Putting this all together, we get

\begin{align}
    z_\mathrm{pt} = \frac{M_{P}}{M_*} \frac{a_{l}}{(a_{l} - a_\mathrm{P} (1+e_P)) }\sqrt{a_{P}a_{l}\frac{(1+e_P)}{(1-e_P)}} (sin (i_P))^{-1}
    \label{particleheight}
\end{align}

\begin{figure}
    \centering
    \includegraphics[width=0.45\textwidth]{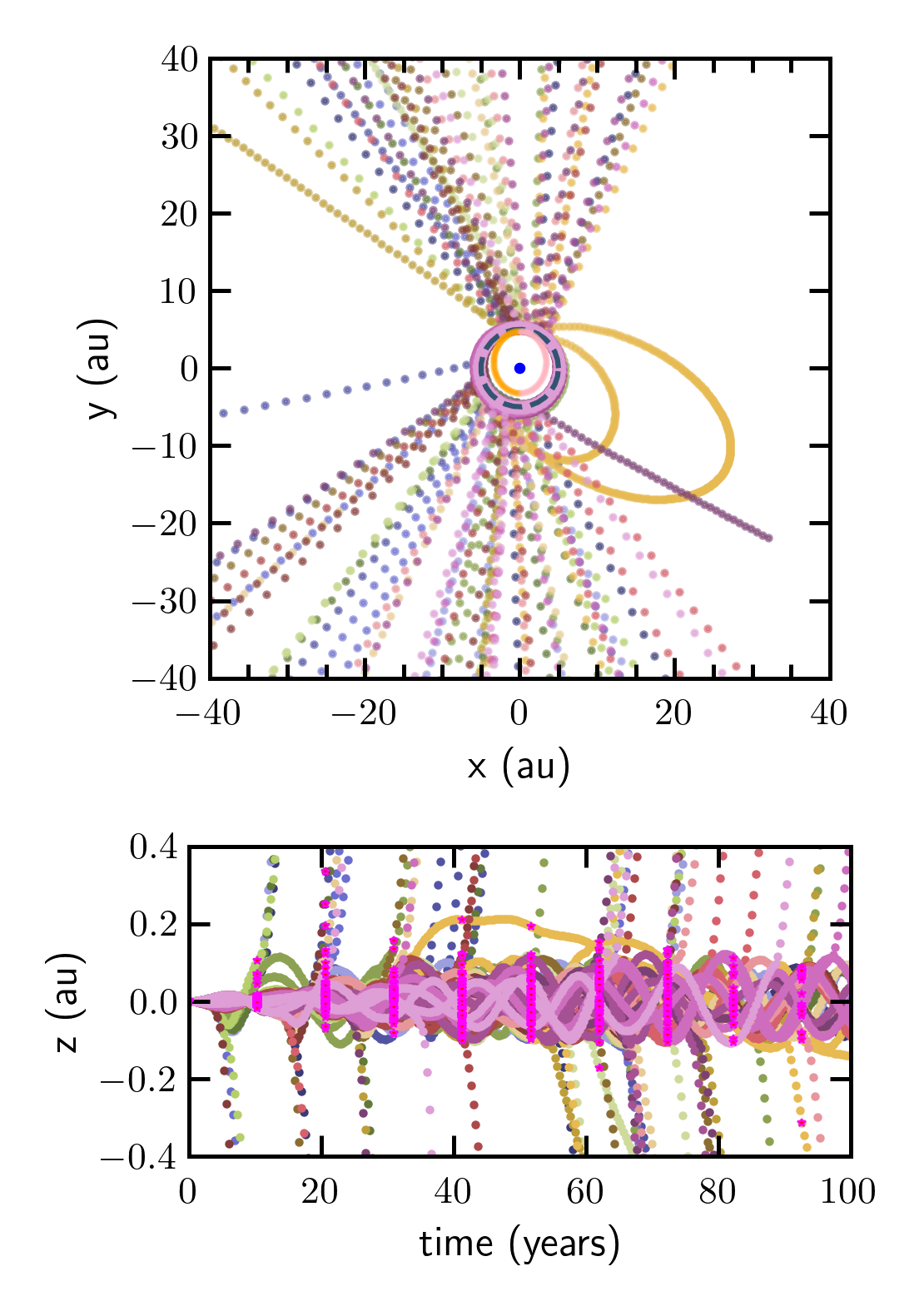}
    \caption{Example of particles' trajectories (top) and height over time (bottom), showing the behavior that produces particle ejections every period. Particles are ejected in the negative and positive vertical direction. They are ejected by the stellar wind when their height reaches a certain value, $z_h$, which is left as a free parameter in the simulation. We choose the highest $z_h$ possible (i.e. a value close to the local minimum/maximum of the particle height over time). This ejection occurs in the orbital region close to the planet's apocenter and pericenter. In this example the planet ascends at apocenter and orbits counterclockwise. The particles launched toward -$y$ are moving in the positive $z$ direction, and those moving toward +$y$ are moving in the negative $z$ direction. This simulation includes a $0.6 M_\odot$ star, a $2 M_J$ planet, and 200 massless particles. The planet is initialized at $a_P = 4$ au, $e_P = 0.3$, $i_P = 30^\circ$. The particles are uniformly distributed between $5.5-6$ au on coplanar, circular orbits.}
    \label{fig:particle_trajectory_height}
\end{figure}

Typically, we find that when $R_d/a_P \lesssim 2$ the particles reach heights that are a factor of $2-3\times$ lower than calculated in Equation (\ref{particleheight}). The simulation results deviate further from the order of magnitude calculation when $R_d/a_P \gtrsim 2$, and so the impulse approximation breaks. At these larger separations, the particle's motion is no longer dominated by the planet's periodic impulses but the averaged planetary motion. With these tests, we confirm that particles will feel periodic kicks when the separation between planet and disk is small and the impulse approximation is valid. 

The force imparted onto the dust will also depend on
the planet's mass, eccentricity, and inclination. 
If the force is too small, there will not be a significant periodic impulse imparted on the particles. If the force is too large, the force may disrupt a larger sphere of influence and hinder periodic behavior. The eccentricity should be high enough so that a larger impulse is felt at apocenter, but not so large to create a crossing orbit which would disrupt the disk. 

Since the observed puffs are believed to have a periodicity of about 7--10 years, the semi-major axis of the planet should be about $3-4$ au. For the planet to kick disk particles every period, the disk's inner edge should then be between 3.5-9 au. Planets with a mass $\gtrsim 3 M_J$ (Jupiter-mass) and distance from the star $\gtrsim 3$ au would have $\gtrsim 50 \%$ probability of being detected with JWST \citep{Lawson2023}, so these parameters are within a region where a planet would not be obviously detectable. Using these values, Equation (\ref{particleheight}), and the height of observed puffs, we place a broad constraint on the planet's mass and run further test simulations. 

We varied the mass of the planet between $1-3 M_J$. We varied the semi-major axis of the planet between $3-5$ au and that of the test particles between $3.5 - 15$ au, where the debris disk inner edge is always exterior to the planet's orbit. We varied eccentricities of the planet between 0 to 0.9$e_{\rm cross}$, where $e_{\rm cross}$ is the minimum eccentricity where the planet's orbit and disk orbit cross (i.e., when apocenter of the planet is equal to or greater to the disk's inner edge, $e_{\rm cross} =  R_d/a_P -1$). We varied the inclination of the planet between 10-60 degrees. The disk particles were uniformly distributed in a ring 0.5 au wide and initialized in coplanar, circular orbits with uniformly randomized phase angles (argument of pericenter ($\omega$), longitude of ascending node ($\Omega$), true anomaly ($f$)) from 0 to 2$\pi$. 

Dominated by the planet's orbital motion, the particles' height over time follows a sinusoidal pattern (Figure \ref{fig:particle_trajectory_height}, bottom). On top of this sinusoidal motion, the planet induces an instantaneous kick which will push some particles above the threshold $z_h$. Remember $z_h$ is the height in the simulation when the particles ``feel'' the stellar wind, and it is defined as $h r$, where $h$ is the disk's aspect ratio and $r$ is the distance of the particles. We find that $z_h$ should be approximately the maximum height the particles reach, so that particles are emitted periodically, as seen in the bottom panel of Figure \ref{fig:particle_trajectory_height}.
Particles are ejected from both the positive and negative direction. Considering our model at face value, we expected the majority of particles to be ejected in the +$z$ direction, but we did not find a preference for release after apocenter passage. This is likely due to particles' orbits becoming similar to the planet's inclined orbit, and particles are released once they pass the height threshold. This threshold could be either after the planet's pericenter or apocenter passage. The last parameter to consider is $\beta$. For given planet parameters and varying $\beta$, we found that decreasing $\beta$ means that the particles will be ejected with their tangential orbital velocity dominating. If $\beta$ is larger, then the particles are emitted in a more radial direction from the star.

In Figure \ref{fig:particle_trajectory_height}, we show an example of particles' trajectories due to a planet's gravitational influence and stellar wind. In this example, $M_P = 2M_J$, $a_P=4$ au, $e_P=0.3$, $i_P = 30^\circ$, $R_d = 5.5$, $\beta = 10$, and $z_h = 0.02r$ au. Particles above the threshold feel the stellar wind and are pushed out radially; this behavior happens periodically. We found that particles are not necessarily ejected at the same location where the planet reaches apocenter; it takes time for the particles to continue on their path before they are far enough from the mid-plane to be ejected. This may take half an orbital period. The planet is orbiting counterclockwise and ascends at apocenter. The particles kicked above the mid-plane are those toward the bottom of the plot, whereas particles kicked below the mid-plane are those towards the top of the plot. 
Given our test simulations, we fix the planet's mass, eccentricity, and inclination to 2 $M_J$, $0.8e_{\rm cross}$, and $30^\circ$, to be used for the large scale simulation. By reducing the allowed parameter space, we then use an Markov Chain Monte Carlo (MCMC) process to constrain the remaining parameters: planet period (and therefore semi-major axis), particle disk inner edge, and stellar wind force. We note that there are a range of other possible parameter combinations and we have not explored all of them in this work. 

\begin{table}
    \centering
    \tabcolsep=0.11cm
    \begin{tabular}{ccccc}
        \multicolumn{5}{c}{prior bounds}\\
        \hline
        $R_d$ (au) & $a_P$ (au) & $P$ (yr) & $\omega_l$ (rad) &$\beta$ \\
        \hline
        $2-30$ & $0.65R_d - R_d$ & $1-20$ & $0 - 2\pi$ &$1-20$\\
        \hline
        \hline
        \multicolumn{5}{c}{best fit parameters}\\
        \hline
        $t_{\rm o}$ (yr) & $R_d$ (au) & P (yr)& $\omega_l$ (rad) &$\beta$ \\
        \hline
        $1.56_{{-{0.24}}}^{+{0.21}}$ & $5.35_{{-{0.09}}}^{+{0.08}}$ & $8.56_{{-{0.15}}}^{+{0.15}}$ & $0.036_{{-{0.10}}}^{+{0.11}}$ & $1.76_{{-{0.18}}}^{+{0.22}}$\\
    \end{tabular}
    \caption{Best fit parameters from the MCMC.}
    \label{tab:MCMC_bestfit}
\end{table}

\begin{figure*}
    \centering
    \includegraphics[width=7in]{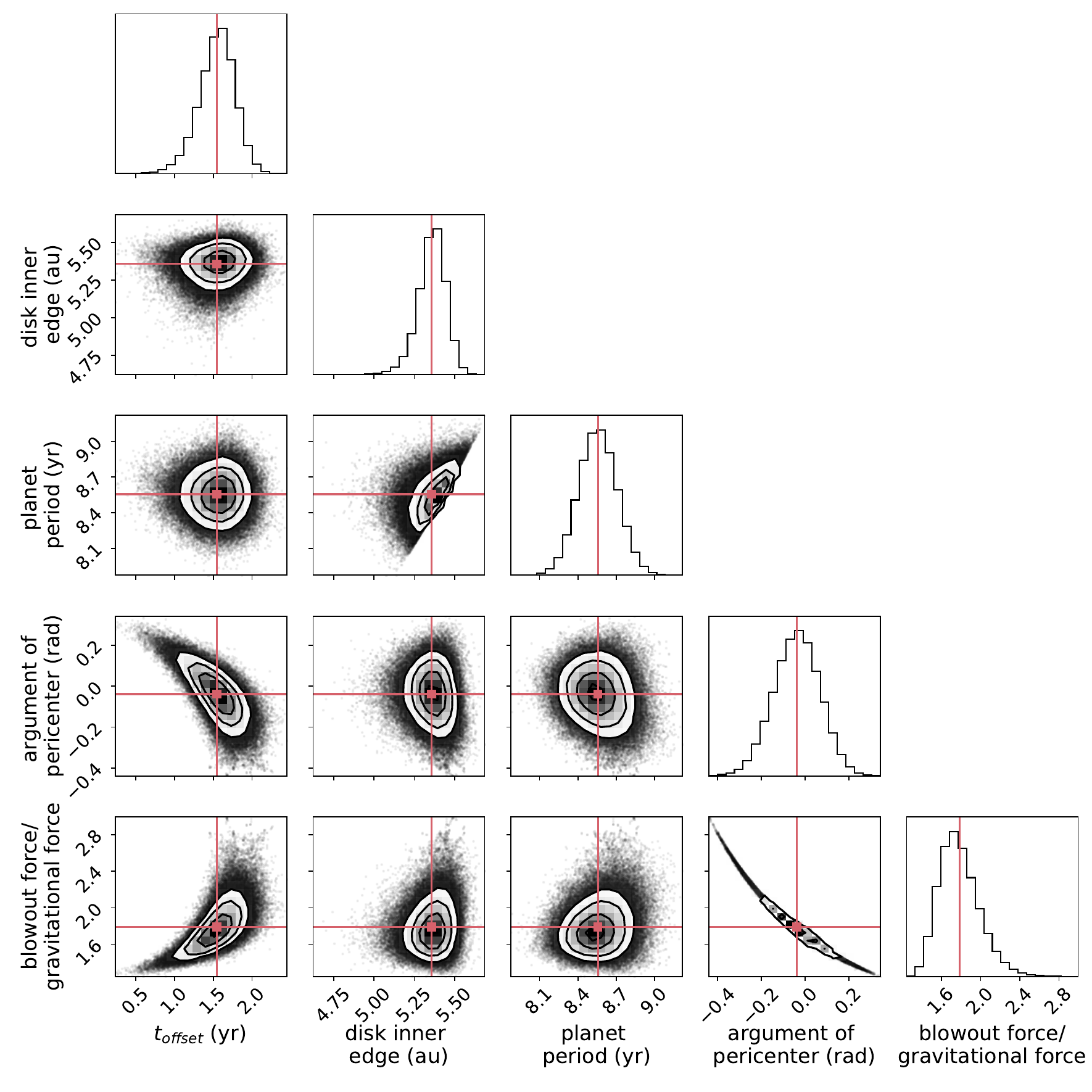}
    \caption{Corner plot showing the best-fit parameters for $R_d$, $P$, $\omega_l$, and $\beta$ to produce particle trajectories that match the projected distance and speed and periodicity of data from \citet{Boccaletti2018} (see, e.g. Figure \ref{fig:velocity_position_data}). The prior bounds and best-fit values are shown in Table \ref{tab:MCMC_bestfit}. The $t_{\rm off}$ parameter is not relevant for the actual simulations, but it helps find a better fit for the period of the planet. The remaining best-fit values are used for a large scale simulation.}
    \label{fig:mcmc}
\end{figure*}

\subsubsection{Particle Trajectory and Velocity Fits with MCMC}
\label{sec:MCMC}

We are specifically interested in finding best-fit parameters for the inner edge of the disk ($R_d$), blowout angle ($\theta$), period of the planet ($P$), stellar wind force ($\beta$), and offset timescale to help match the periodicity between the planet and the clumps ($t_{\rm off}$\footnote{$t_{\rm off}$ is the time between the beginning of the simulation and when the particle trajectory reaches the first clump in Figure \ref{fig:velocity_position_data} (i.e., time before the first filled circle). This parameter is not relevant in simulations. It simply helps fit a planet orbital period to the data.}). We use the MCMC formulation from \citet{Goodman2010} implemented as the Python package \texttt{emcee} \citep{ForemanMackey2013}.
As is commonly the case, we used a log-likelihood metric, $\mathcal{L}=-\chi^2/2$, to determine the quality of fit between the particle trajectories and data.

We assumed uniform priors for all parameters with bounds shown in Table \ref{tab:MCMC_bestfit}.
We performed MCMC runs for 5 parameters with 100 walkers with a burn-in of 1508 and sampled 10$^6$ times. For each run, a new instance of a simulation with a single particle runs for various combinations of ($R_d$, $\theta$, $\beta$). Then, the routine calculates how well the particle trajectory fits the data in Figure \ref{fig:velocity_position_data}, and finds a best-fit planet period that fits the periodicity of the data. The particles are initialized with $e=0$, $i=0$, $\Omega = 0$, $f = 0$. The argument of pericenter, $\omega$ is the same as the blowout angle in this case.
The best-fit parameters are shown in Table \ref{tab:MCMC_bestfit}. In Figure \ref{fig:mcmc} we show the corner plot for the MCMC made with \texttt{cornerplot}, showing contours for the 0.5, 1, 1.5, and 2 sigma confidence regions for a 2D Gaussian distribution.

\section{Large Scale Simulation Results}
\label{sec:large_scale_simulation}

With our pilot tests, we gained intuition regarding the dynamics of the system and constrained parameter space for orbital configurations that produce periodic puffs with velocities that are comparable to the observed values. Here, we present results for the large-scale simulation and visualize the debris disk to compare qualitatively and quantitatively with observations.

\subsection{Simulation Setup}
\label{sec:large_sim_setup}

\begin{figure}
    \centering
    \includegraphics[width=0.45\textwidth]{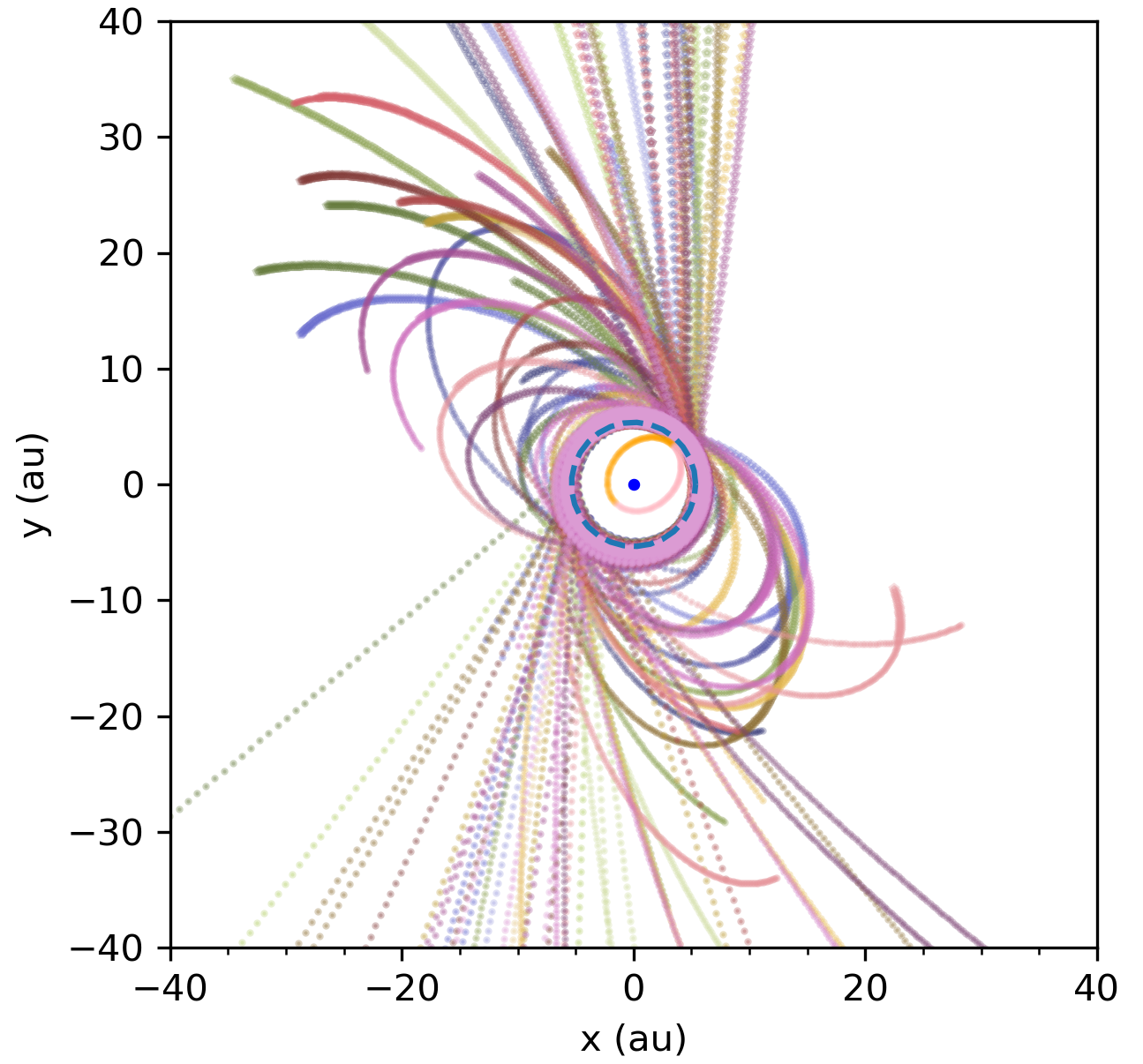}
    \caption{Example particle behavior for a simulation with a $0.6 M_\odot$ star, $2 M_J$ planet, and 200 massless particles. The planet is initialized with $a_P = 3.52$ au, $e_P = 0.37$, $i_P = 30^\circ$. The particles are uniformly distributed between $a_l = 5.36 - 5.86$ au on coplanar, circular orbits. $\beta = 1.76$. The low $\beta$ value means particles feel a smaller outward acceleration compared to the simulation in Figure \ref{fig:particle_trajectory_height}, where $\beta = 10$. Since the outward acceleration is low,} particles near $z_h$, might turn off abruptly. We do not expect the particles on eccentric orbits as shown here to be physically meaningful since there would not be an abrupt stop to the stellar wind in reality.
    \label{fig:particles_trajectories_beta1.7}
\end{figure}

In a few more test simulations with a planet at $3.52$ au and disk between $5.36-5.86$ au with $\beta = 1.76$, we find that many particles linger near $z_h$ and feel the stellar wind for a short time, however they then fall below the threshold and the stellar wind turns off. This pushes particles radially outward but not into inclined hyperbolic orbits; instead, they remain close to the mid-plane on elliptical orbits, as seen in Figure \ref{fig:particles_trajectories_beta1.7}. Note the difference with Figure \ref{fig:particle_trajectory_height}, which had $\beta=5$. Viewed by an observer, we expect the particles on elliptical orbits to remain near the main disk and be completely separate from the clumps. In this section we focus on analyzing the clumps. In future work it will be interesting to investigate whether the particles on elliptical orbits create a dynamically hot population that is distinguishable from the main disk.

We confirm that the small-scale simulations produce disk morphologies and clump velocities that match the observations reasonably well, and so we move on to run a large-scale simulation. In the large scale simulation, the planet is initialized with mass ($M_P = 2 M_J$), semi-major axis ($a_P = 3.52$ au), eccentricity ($e_P = 0.37$), inclination ($i_P = 30^\circ$ ), longitude of ascending node ($\Omega = 0$), argument of pericenter ($\omega = \pi$), and true anomaly ($f = 0$). The debris disk particles are uniformly distributed between $R_d = 5.36$ au and $R_{out} = 5.8$ au on circular, coplanar orbits. The stellar wind force $\beta$ is 1.76 for particles above z threshold ($z_{h} = hr$), where $h=0.023$ is the aspect ratio and $r$ is radial distance from the star. In Section \ref{sec:discussion}, we discuss the best-fit value of $\beta$ as it relates to AU Mic stellar wind velocities and mass loss rates from the literature. If a particle drops below $z_h$, $\beta$ is 0. This condition is checked at every timestep (0.005 yrs). The total simulation runtime is 200 years. For computational efficiency, we split the simulation of 240,000,000 particles across 240 instances of the same simulation with 100,000 randomized particles in each. Across the 240 simulations, the star and planet evolve the same way, and the test particles have a random seed attributed to them. 

\begin{figure*}
    \centering
    \includegraphics[width=\textwidth]{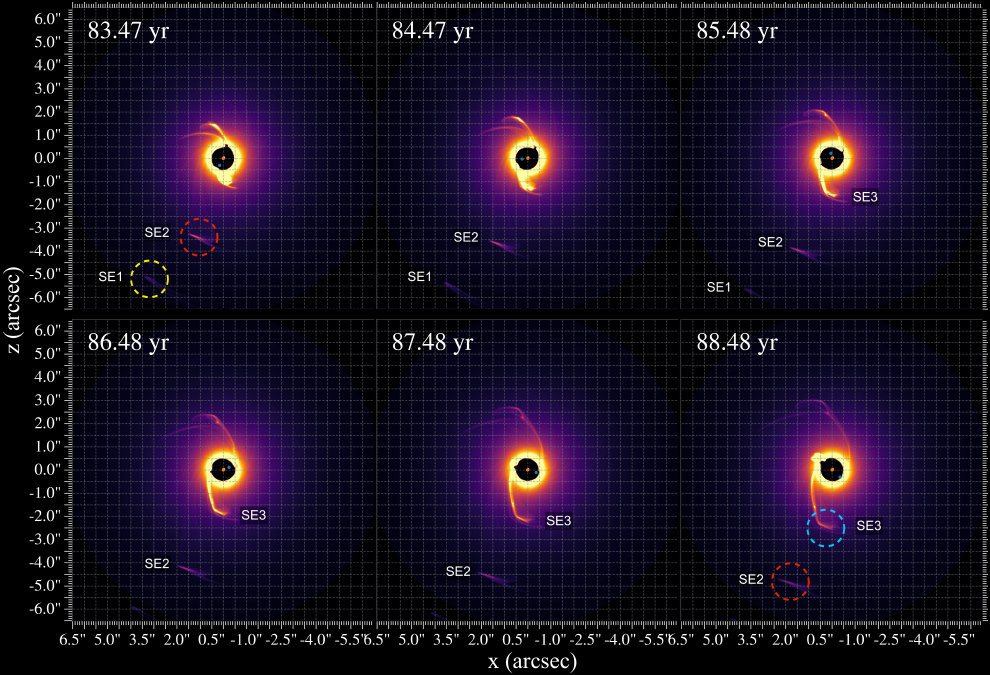}
    \caption{Snapshots, one year apart, of the simulation described in Section \ref{sec:large_scale_simulation}, as seen from above. Yellow, red, and blue circles with a dashed outline are shown to denote 3 different particle ejections. The same circles show up in Figure \ref{fig:disk_edgeon} to connect features in the top-down view with features seen from an edge-on perspective.}
    \label{fig:disk_topdown}
\end{figure*}

\begin{figure*}
    \centering
    \begin{subfigure}{3.5in}
        \includegraphics[width=3.5in]{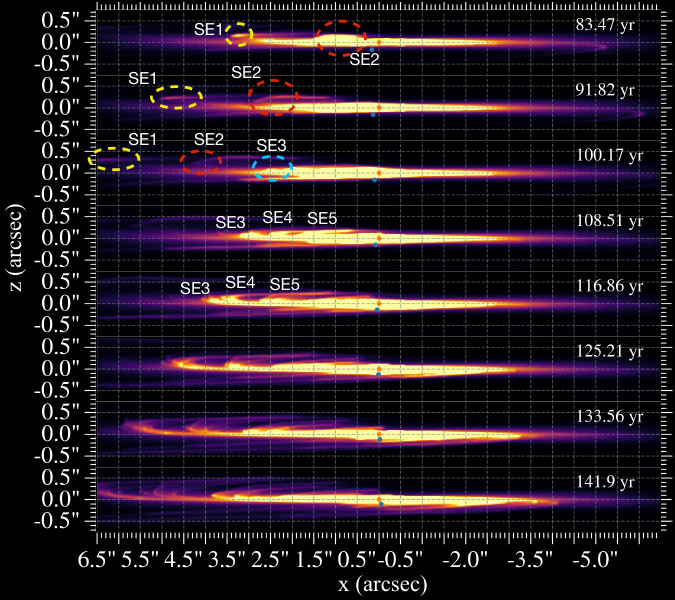}
    \end{subfigure}
    \begin{subfigure}{3.5in}
        \includegraphics[width=3.5in]{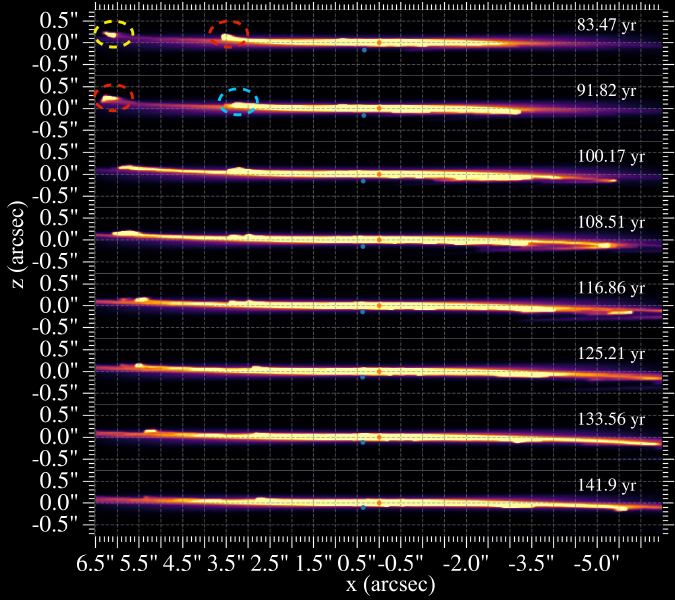}
    \end{subfigure}
    \caption{Edge-on view of the disk and clumps every 8 years for two different observer lines of sight. Dust clumps on the top, left side of the disk are denoted with yellow, red, and blue circles, which correspond with circles shown in Figure \ref{fig:disk_topdown}. \textit{Left:} The observer line of sight is at 6 o'clock in Figure \ref{fig:disk_topdown}. In these frames, puffs only occur on the left side. New clumps build up every $\approx 8-9$ years. The clumps become elongated and the tail gets very dim. Looking at the top-down image, this behavior is noticeable in following the path of the bar feature. \textit{Right:} The observer line of sight is at $\approx 4$ o'clock in the top-down image. The clumps appear as smaller features and are bright for the majority of their trajectory. At this angle, the particles going in the $-z$ direction move to the right of the disk, moving outward. This behavior does not match observations. The puffs have a larger projected speed in this configuration. }
    \label{fig:disk_edgeon}
\end{figure*}

\subsection{Visualizing the Disk}
\label{sec:visualize_disk}

We visualize the simulation snapshots with \texttt{numpy.histogram2d}, which bins the particle positions into 2D pixels and weights the surface brightness with a scaling of $r^{-0.5}$. The flux of the disk is normalized by an arbitrary value, and a central flux representing the star is added. This type of surface brightness scaling is typical for thermal emission and used to create synthetic observations to compare to ALMA images. The clumps in the AU Mic disk are only detected in scattered light images, and directly comparing simulations to those images requires passing the simulation output through a radiative transfer code such as \texttt{RADMC}, which we leave for future work. We do not intend to analyze the relative brightness of the clumps and disk, since that will differ in scattered light images. Instead, we choose to visualize the simulated disk with surface brightness maps to analyze the general morphology, position, and velocity of the clumps and compare them to the observations.

The large-scale simulation described in Section \ref{sec:large_sim_setup} contains particles between 5.36--5.86 au. For the surface brightness maps, we superimpose a static disk from 5.86 au to 60 au to represent the larger debris disk. This disk is created with $10^8$ randomized points representing a circular, co-planar disk. The 2D array is generated in the same way as described above. Since the surface density of the two disks is different, we scale the 2D array of the simulated disk down by eye. Then the two arrays are superimposed to create the surface brightness maps shown in this section. The $x-y$ and $x-z$ maps have dimensions 13 arcsec by 13 arcsec and 1.5 arcsec by 13 arcsec, respectively. 

\subsubsection{Comparison to Observations}

When searching for an edge-on perspective that had the best qualitative match to the observations of AU Mic, we found that rotating the disk 180 degrees about the x-axis and 60 degrees about the z-axis produced the best match. This means that the planet and grains from this point of view are orbiting clockwise. Figure \ref{fig:disk_topdown} shows the top-down view of our simulation, where each panel represents snapshots one year apart. These surface brightness maps clearly show two puffs that were ejected at an earlier time (SE1'; yellow circle and SE2'; red circle) and a new puff (SE3'; blue circle). We use the same labels as \citet{Boccaletti2018} to make connections with their observations and add an (') in the text when referencing the simulation clump. Due to the rotation, the planet descends at apocenter and ascends at pericenter in these images. The coherent groups of dust particles moving downward and upward in Figure \ref{fig:disk_topdown} are emitted in the positive $z$ direction (out of the plane) and negative $z$ direction (into the plane), respectively.

The edge-on surface brightness maps are most helpful for comparing the morphology of our simulations with the observations.
Figure \ref{fig:disk_edgeon} shows the edge-on perspective of the disk from two different observer lines of sight, which are at 6 o' clock (left panel) and 4 o'clock (right panel) when looking at the top-down view in Figure \ref{fig:disk_topdown}. The first snapshot in Figure \ref{fig:disk_edgeon} is taken at the same time as the first snapshot in Figure \ref{fig:disk_topdown} and the remaining panels represent snapshots taken $8-9$ years apart. As expected by our pilot simulations, a new puff is emitted every planetary orbital period (8.5 yrs). The configuration on the left panels produces the majority of ejections at the top left of the disk and proves to be the best qualitative match to observations of AU Mic. Each individual puff is labeled SE1' (yellow oval), SE2' (red oval), SE3' (blue oval), SE4', and SE5' to make connections with the top-down configuration and the observations from \cite{Boccaletti2018}. The configuration on the right also contains the ovals to differentiate between the different puffs. From this perspective, the clumps look smaller, the disk is inclined, and several ejections also occur on the bottom right. Unlike the perspective on the left, the perspective on the right does not provide a good fit to the data. See Appendix \ref{sec:appendix} for a visualization with more viewing angles and an accompanying movie in the online version.

\begin{figure*}
    \centering
    \includegraphics[width=\linewidth]{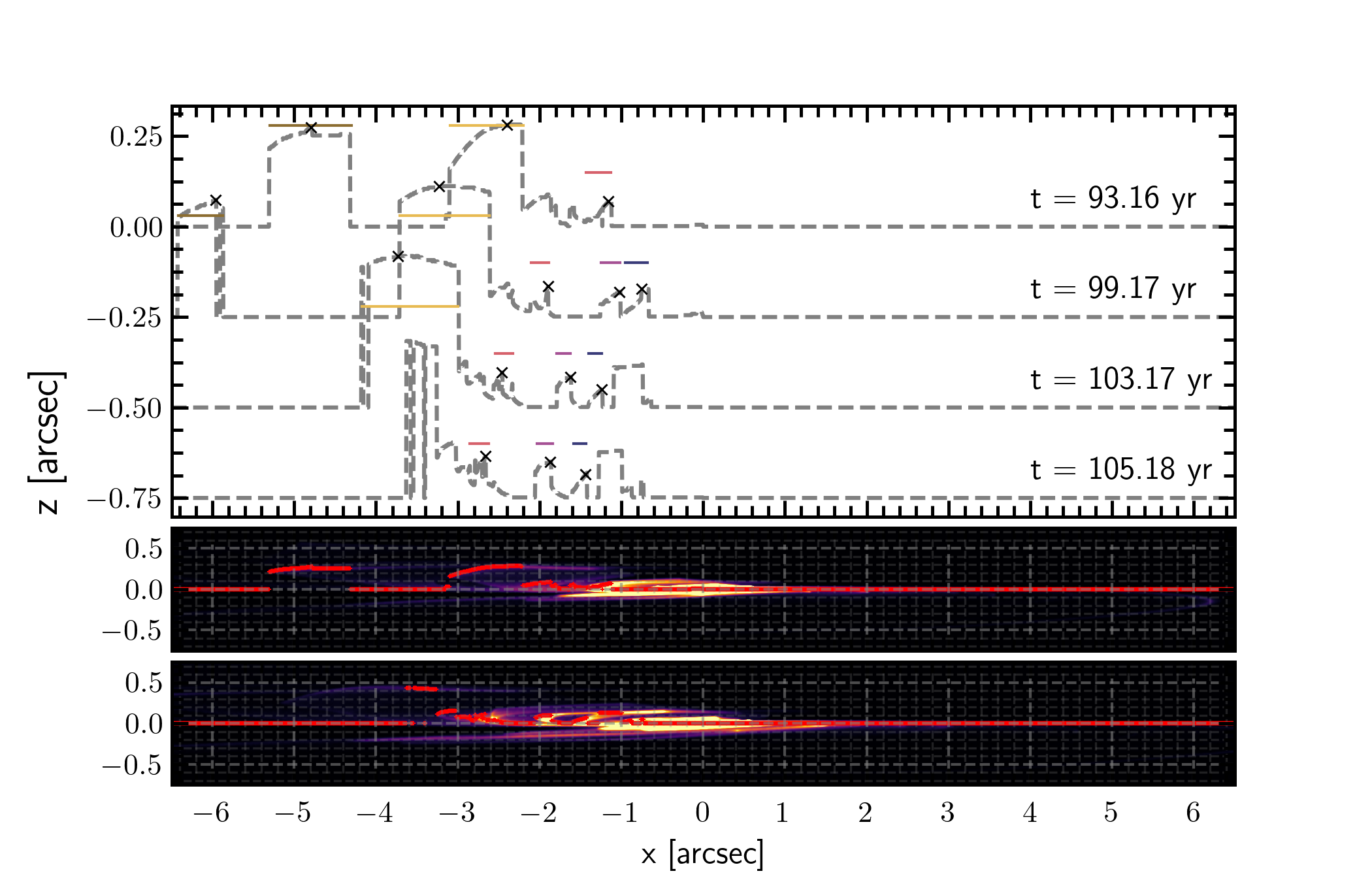}
    \caption{ The ''spine" of the simulated disk at four different time snapshots. This is calculated by finding the pixel above the mid-plane and on the left side of the disk where the intensity is highest. The blue, purple, pink, yellow, and brown horizontal lines represent the width and uncertainty in the position of the clumps SE5', SE4', SE3', SE2', and SE1', respectively. The `x' marks the location of the puff, which is where the pixel with the highest intensity for each puff lies. If no line is present, the clump was not clearly visible in the surface density maps. Two surface density maps with the superimposed spine (red) are shown for reference. Note, the extended disk is not included here, so that we could more accurately determine the presence of clumps.}
    \label{fig:disk-spine}
\end{figure*}

Apart from being primarily emitted on the top left side of the disk, the simulated puffs also decrease in brightness and elongate over time, and become spiral-like (see Figure \ref{fig:disk_edgeon} left panel). This is in agreement with observations. They reach elevations between 0.5 and 0.15 arcsec (Figure \ref{fig:disk-spine}), which provides a very good match to the observed elevations, and can be as high as 0.3 arcsec at $\sim$6 arcsec radial separation from the star. The arch-like features seen in the edge-on visualizations are the bar-like features circled in Figure \ref{fig:disk_topdown}. This arch shape is also seen in the observations–see specifically SE3 from Figure 6 of \citet{Boccaletti2018}. 
In future work, we will compare synthetic scattered light images to observations and will be able to comment further on relative brightness asymmetries.

We calculate the ``spine'' of the disk following a similar methodology as discussed in \citet{Boccaletti2018}. We measure a height above the mid-plane (i.e, above 0.081 arcsec) where the intensity is highest. We check the intensity one pixel above the measured height, and if it is 0.9 of the max intensity, then we choose that pixel instead to get a more representative highest point of the clump. We calculate the spine for four snapshots that are 6, 4, and 2 years apart, relative to each other. To measure the spine, we use surface brightness maps with a higher resolution that do not include the extended disk so that the clumps are more visible and we can more easily track their evolution. Figure \ref{fig:disk-spine} shows the temporal evolution of the spine and two disk surface brightness maps with the spine superimposed for reference. 

\begin{figure}
    \centering
    \includegraphics[width=\linewidth]{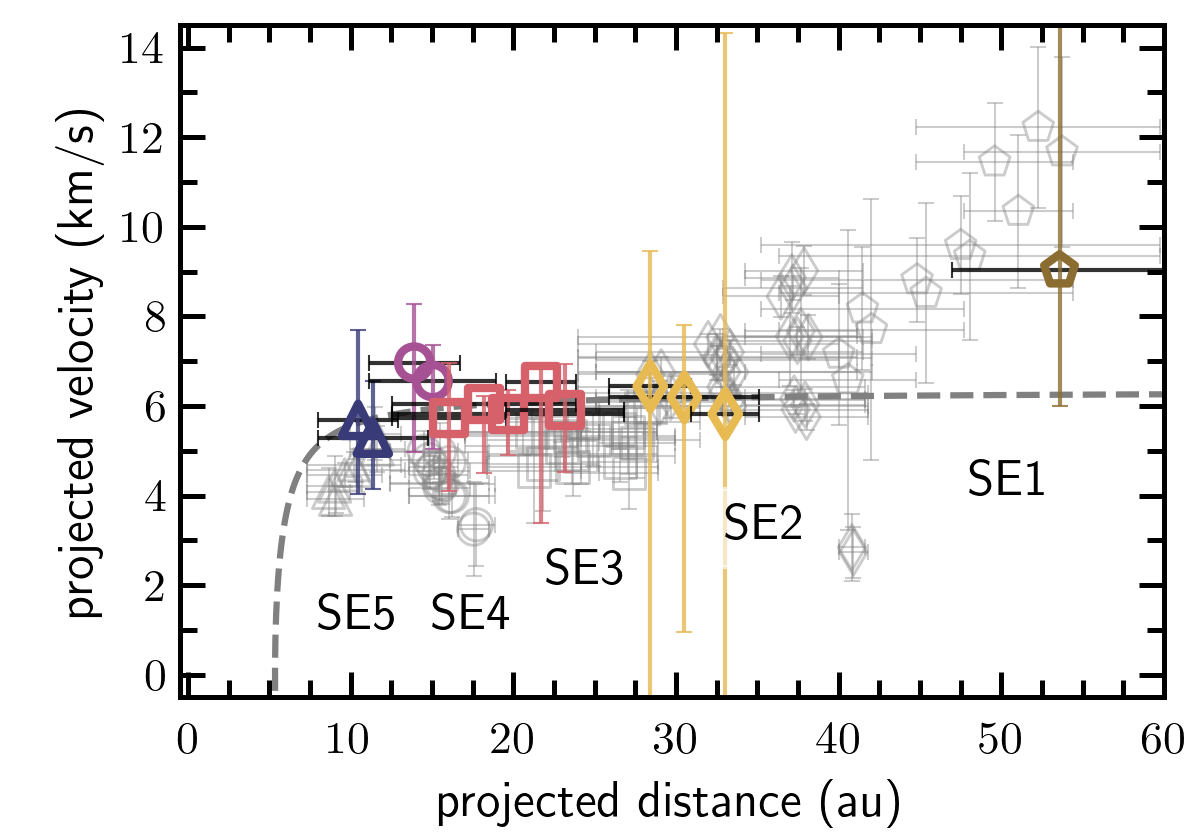}
    \caption{Projected velocity vs projected distance of the 5 clumps in our simulation: SE5' (triangle), SE4' (circle), SE3' (square), SE2' (diamond), SE1' (pentagon). The velocities are calculated relative to the positions at 93.16 yr and 99.17 yr (see Figure \ref{fig:disk-spine}). The black horizontal lines represent the temporal baseline between two epochs for which the projected velocity was evaluated (i.e., the clump's position at each epoch was at either end of the line). The vertical error bars represent the uncertainty in the positions of the clumps. The projected velocities for the features found in the AU Mic images are shown in grey. The grey dashed line represents the expected particle trajectory based on our MCMC and large-scale simulation. We note that at large projected distances, the clumps become azimuthally extended, and the location of their projected maximum brightness deviates from a single-particle trajectory, consistent with AU Mic observations. }
    \label{fig:velocity-compare}
\end{figure}

Next, we calculate the projected speeds of the five clumps. We measure the position and correlated uncertainty of each clump to be the location with the highest intensity and the remaining width of the clump, respectively (see top panel of Figure \ref{fig:disk-spine}). We calculate the velocity of each clump relative to the positions at 93.16 yr and 99.17 yr, where the temporal baselines between epochs are approximately the same as those in \citet{Boccaletti2018}. The velocities of the simulated clumps are shown in Figure \ref{fig:velocity-compare}, where the clumps are differentiated by color and shape. Our calculated velocities are in agreement with the AU Mic data (grey shapes), within the error bars. The grey dashed line represents the expected velocity trajectory of the particles in our simulation, based on the results of the MCMC (Figure \ref{fig:mcmc}). We confirm that the simulated puffs actually follow this trajectory. We expected the velocity of simulated clump SE1' (brown pentagon) to sit closer to the curve as well. Interestingly, the velocity is higher than the rest and therefore produces a better match to the data.  We note that there are larger positional uncertainties for clumps farther from the star since they spread out over time. Also, tracking the pixel with the highest intensity may not be tracking the position of the same particles over time. It is possible that this limitation is also affecting the measured velocities inferred from observations. The velocities measured from outermost observed clump may be overestimated relative to what would be obtained by tracking a single grain over time. Something to note is that in our simulations we find cases where particles belonging to newly formed clumps surpass older clumps in projected distance. This is especially noticeable in snapshots between 110-135 years (see Figure \ref{fig:disk_edgeon} and panel B in the animated Figure \ref{fig:disk_diff_angles}). This may suggest that high intensity pixels in the furthest clump are not representative of locations for particles that were originally emitted together.
Nonetheless, our results produce promising agreement with the observations and merit follow-up analysis when future observations are available.

\section{Discussion}
\label{sec:discussion}

Here, we discuss our assumptions, existing constraints from observations, and considerations for future work. 

\subsection{Stellar properties and grain size}\label{sec:stellarprop}

For the planet-disk model presented in this paper, we find the best match to observations when $\beta = 1.76_{{-{0.2}}}^{{+0.2}}$, based on the MCMC result and simulation. Since radiation pressure is not expected to blow out grains of any size in this system \citep{Arnold2019}, $\beta$ is only the ratio of stellar wind force and gravitational force:

\begin{equation}\label{eqn:beta}
    \beta = \frac{3}{32\pi}\frac{\dot{M_*}V_{sw}C_D}{GM_*\rho s}
\end{equation}

\noindent where $\dot{M_*}$ is the stellar mass loss rate, $V_{sw}$ is the solar wind speed, $C_D$ is a drag coefficient (2), $G$ is the gravitational constant, $M_*$ is the mass of the star ($0.6 M_\odot$), $\rho$ is the grain density, and $s$ is the size of the grain.

The relation between $\beta$, $\dot{M_*}$, $V_{sw}$, and $s$ is convoluted and here we make sense of how $\beta = 1.76$ fits in with values quoted in the literature. \citet{Arnold2022} explored how the dust grain shape affects the composition and grain size distribution of the AU Mic debris disk using scattered light data. They find that a smallest grain size of 0.2 $\mu m$ provides the best fit to the data. There are large degeneracies within the compositions for spherical and agglomerate shapes, though silicates appear to be the most abundant. We use $\rho=1.78 g/cm^3$ which corresponds to a composition of silicates, carbon, and vacuum equally distributed \citep{schuppler2015}. The values for stellar mass loss rate are variable due to the nature of M stars; values in the literature are as low as 10 $\dot{M_\odot}$ and up to 2500 $\dot{M_\odot}$ \citep{Strubbe2006,Plavchan2009,Augereau2006}, where $\dot{M_\odot} = 1.265 \times 10^{12} g/s$. \citet{AlvaradoGomez2022} found stellar wind velocities between 1150 and 2200 km/s, though their models yielded lower values of mass loss rate between $5-10 \dot{M_\odot}$. We adopt $\dot{M_*}=300 \dot{M_\odot}$ and $V_{sw}= 2200$ km/s. Coincidentally, putting this all together leads to $\beta=1.76$, matching our best fit value. This confirms that our model is consistent with a set of reasonable physical parameters.

\subsection{Height Dependence of Stellar Wind Strength}\label{sec:height_beta}

In the simulations presented in this paper, the stellar wind is turned on for particles whose height surpasses a threshold value. This choice is motivated by three different scenarios (Section \ref{sec:planet_disk_model}) that may enable radially accelerating dust grains above the mid-plane while the mid-plane material appears mostly undisturbed. Here, we present order of magnitude estimates to assess how plausible each scenario is.

We expect a reservoir of dust to remain at all times due to collisional processes between large grains and planetesimals. ALMA observations and subsequent modeling support the presence of an inner ring at $<10$ au \citep{MacGregor2013, Daley2019, Han2026}. \citet{MacGregor2013} constrained the inner component of AU Mic to have $\sim 10^{-4} M_\oplus$ of mm grains, while the outer belt centered around 40 au has 0.01 $M_\oplus$ in mm grains. Through collisional modeling which included stellar winds, \citet{schuppler2015} estimated a minimum disk mass for grains extending up to large planetesimal sizes to be 0.02 $M_\oplus$ in the inner region. While these values hinge on assumptions for grain opacities and temperatures, they provide order of magnitude estimates for reference.

In the first scenario, we posit a large dust density in the midplane so that some grains are shielded from direct encounters from the stellar wind, resulting in particles above the midplane feeling the stellar wind more strongly.
Extrapolating from the mass estimate for mm grains from ALMA observations and assuming a size distribution with power-law index q = -3.5, we might expect $M_{0.2\mu\text{m}}=10^{-4} \big(\frac{0.2 \mu m}{1 \rm{mm}}\big)^{0.5} \sim 1.4 \times 10^{-6} M_\oplus$ in submicron-sized grains in the inner disk near our proposed planet.  Considering a fiducial disk with constant surface density extending from 5 to 10 au from the host star and with aspect ratio $h = 0.02$ \citep{Vizgan2022}, this mass yields a midplane volumetric density of $\rho_{0.2\mu\text{m}}\sim5\times 10^{-20}$ g cm$^{-3}$ at $r=5$ au in 0.2-micron grains. Using the fiducial stellar wind mass loss rate, $\dot M_*$, and velocity, $V_{sw}$, described in Section \ref{sec:stellarprop}, the volumetric mass density of stellar wind plasma at the same location is substantially smaller: $\rho_{\rm sw} = \dot M_*/(4\pi r^2 V_{sw}) \sim 2.5\times 10^{-23}$ g cm$^{-3}$.  Under such conditions, entraining disk grains will substantially slow stellar wind particles.  If the grain density is large enough that all wind particles are expected to collide with grains, the effective $\beta$ provided by the combined outflow of grains and stellar wind plasma in the midplane (c.f. Equation \ref{eqn:beta}) will be substantially reduced.  

Such self-shielding requires the radial optical depth of grains with radius $s = 0.2\mu$m (equivalent to the stellar wind particle encounter probability, assuming a geometric cross section, $\sigma = \pi s^2$) in the midplane to be greater than 1, so that some grains are shielded from direct encounters with stellar wind plasma. The radial optical depth over the $\Delta r = 5$au radial extent of our fiducial inner disk is $\tau \sim (\rho_{0.2\mu\text{m}}/m)\sigma\Delta r \sim (\rho_{0.2\mu\text{m}}/\rho)(\Delta r/s) \sim 8\times 10^{-2} (0.02/h)(M_{0.2\mu\text{m}}/(1.4 \times 10^{-6}M_\oplus))$, where a grain's mass is $m = (4/3)\pi \rho s^3$ with internal density $\rho = 1.78$ g cm$^{-3}$ (see Section \ref{sec:stellarprop}).  Our fiducial population of submicron particles does not yield significant self-shielding.  However, some models of disk observations allow for values of $h$ as small as 0.0009 \citep{Zawadzki2026} and the mass in submicron grains is not directly constrained, so midplane self-shielding is not entirely ruled out. 

We comment that, though not favored by our fiducial disk, midplane self-shielding merits continued consideration because \citet{Vizgan2022} observed the AU Mic disk to be thicker at 1.3 mm ($h= 0.025$) than at 450 µm ($h = 0.019$).  Typically, smaller grains, which dominate shorter-wavelength observations, are dynamically excited more easily and are thus expected to have larger scale heights.  Stellar wind shielding in the midplane could help explain this puzzling result.  Stellar wind $\beta$ scales as the inverse of particle size (c.f. Equation \ref{eqn:beta}), so particles of radius 450$\mu$m are expected to experience a factor of $\sim$2000 smaller acceleration by the stellar wind than particles of radius 0.2$\mu$m, yielding $\beta \sim 10^{-3}$, insufficient to blow 450$\mu$m-particles out of the disk. The orbital decay timescale for these particles due to stellar wind corpuscular drag \citep{Burns1979, Strubbe2006} is $t_{\rm corp} = (4\pi\rho s r^2)/(3\dot M_*) \sim 10^5$yr at 5au, shorter than the lifetime of the disk but longer than the expected collisional destruction time.  However, if 450-micron particles in the midplane are shielded from the stellar wind while those above the midplane are not, the particles above the midplane may be differentially ``sandblasted" by sub-micron particles that are being ejected by the stellar wind.  The escape velocity at 5.5au from the star is 14km/s.  At such high velocities, a substantial flux of sub-micron particles can collisionally erode their larger neighbors.  In a high velocity collision, the mass lost from the larger particle in units of the incident mass is referred to as the yield $Y = Y_0(u/u_0)^2$ \citep[e.g][]{Cuzzi1990}.  This parameter encapsulates the ratio between incident collision energy per mass at collision velocity $u$ and the larger particle's binding energy per mass.  Likely originating as fragments in a collisional cascade, 450-$\mu$m grains need not be weak, fluffy aggregates.  The yield parameter for cratering is material-dependent \citep[e.g.][]{Koschny2001}, and exploring a full range of material compositions is beyond the scope of this work.  We merely note that for a value of $Y \approx 10^2$ at $u = 14$ km/s, modest in the context of experimental collisions with strength-dominated bodies, destruction is possible.  Our fiducial $\rho_{0.2\mu\text{m}}$, moving at the escape velocity from the star, will destroy a grain with radius 450$\mu$m on a timescale of only 6 years.  The observed lower scale height observed in 450$\mu$m observations may then result from collisional destruction by a flux of sub-micron particles accelerated by the stellar wind at heights above the midplane.

An alternative possibility is that all sub$\mu$m-sized particles experience the same stellar wind force independent of their height above the disk's mid-plane. Although dust in the mid-plane is continuously driven radially outward, this motion is not directly apparent because the overall flat disk morphology remains unchanged over time. To generate enough dust to surpass the mass of clumps, we need the collision rate to be at least the mass of the clump per ejection timescale. \citet{Chiang2017} calculate the mass of a clump by taking the brightness of clump SE4\footnote{Called cloud B in \citet{Chiang2017}.} to estimate the line-of-sight column density of grains. Together with the projected area of the clump, this gives $M_{\rm{clump}} = 4 \times 10^{-7} \big(\frac{a}{35 au}\big)^2=8\times 10^{-9}$ M$_\oplus$, where $a = 5$ au, the distance of the clumps from the star. A disk of $0.02 M_\oplus$ would be drained in 10 Myr to 20 Myr, providing a marginal result for this scenario.

A third possibility is that each planetary passage triggers a cascade of planetesimal collisions that produce the sub$\mu$m grains subsequently ejected as a clump. To estimate whether this mechanism can generate sufficient dust, we consider a fiducial system consisting of a disk extending from 5 au to 10 au containing $1\times 10^{-4}$ M$_\oplus$ in mm-sized grains, a 2$M_{jup}$ planet at 3 au ($b=2$ au from the inner disk edge) with eccentricity 0.3 and inclination 30$^\circ$, and a $0.6$ M$_\odot$ star. The planet passes the disk with velocity $v_P = \sqrt{(GM_*/a_P)\frac{1-0.3}{1+0.3}}\sin(30^\circ)$, imparting a random velocity $v_{\rm rand} = GM_P/bv_{P}$. to the particles through the impulse approximation. This velocity sets the disk scale height, $H = v_{\rm rand}/\Omega$, where $\Omega = \sqrt{G M_*/R^3}$, and therefore the disk density, $\rho_{mm} =  M_{\rm disk}/(2\pi R_{in} \Delta R H)$. The grain's mass is $m=(4/3)\pi \rho s^3$, giving a number density $n = \rho_{mm}/m$. The number of collisions within the perturbed region is $n^2 \sigma v_{\rm rand} \Delta t V$, where $\sigma = \pi s^2$ is the geometric cross section, $\Delta t =2 b/v_P$ is the interaction time, and $V=(2b)^2 H$ is the interaction volume. This yields $\sim 10^{21}$ collisions, corresponding to the production of $\sim 10^{-9}$ M$_\oplus$ in small grains. Although this is below the estimated mass of the clump, it is within an order of magnitude. Increasing the disk surface density by a factor of 3 would produce sufficient mass in small grains, suggesting that collisional production during planetary passage is plausible.

\subsection{Inner edge of the disk}
To generate periodic particle ejections, we found that an undetected planet orbital period between 8-9 years and an inner edge of the disk between 5-6 au works best. This inner edge value is within the constraints placed by ALMA observations \citep{Hughes2018,Daley2019,Vizgan2022,Han2026}. Future observations of the disk with higher angular resolution can test this immediately. If the inner edge is confirmed to lie this close to the star, it further motivates the model we propose here. On the other hand, if small grains are not found at these radial distances, this model would require revision. 

\subsection{Planet parameters}
Our final planet values for our simulation are $M_P = 2 M_J$, $a_P = 3.52$ au $e_P = 0.37$, and $i_P = 30^\circ$. A high eccentricity and inclination suggests that planet-planet scattering may have occurred, a scenario motivated by the possible misalignment of AU Mic c \citep{Yu2025}. As described previously, we have chosen only one planetary mass, eccentricity, and inclination to explore in detail. While chosen strategically to create a disk morphology that matches observations, these values are not required for this model, and in fact a planet with smaller mass may be preferred in order to maintain the mutual inclination between the planet and the disk without appealing to recent planetary scattering.

A massive planet on an inclined, eccentric orbit residing near an axisymmetric disk would be expected to excite the disk's eccentricity and inclination as the planet tries to drag the disk into the planet's plane \citep{Faramaz2014, Pearce2014, Poblete2023, Sefilian2025}. The planet is expected to shape the disk over a secular timescale, which depends on $a_P/R_d$, $M_P/M_*$, and $P_P$. Using Equations 6 and 7 from \citet{Costa2024}, the secular timescale for a 2$M_J$ planet at 3.5 au perturbing a disk inner edge at 5.5 au is $\approx 6000$ years. In its simplest form, this means that our model of a thin, unwarped disk implies that the planet became eccentric and inclined within the last $\sim$$10^4$ years, much shorter than the age of the system. Increasing the distance of the planet and inner edge of the disk can increase the secular timescale, but in our model, we require the planet to be at around 3.5 au and inner edge at around 5.5 to eject puffs at the correct periodicity and for the planet to kick the dust grains enough. This uncomfortable coincidence may be avoided by considering a less massive planet. For example, 0.01 $M_J$ planet would have a secular timescale of order $10^6$ years. Lower-mass planets can produce  particle trajectories similar to those illustrated in this paper. The kicks will be weaker, and the particles will reach lower elevations from the mid-plane, but we would still expect periodic ejections and velocities comparable to those shown in this work. We found that clumps increase in elevation over time (Figure \ref{fig:disk_edgeon}); if a lower-mass planet were responsible for these clumps, this model still works if the clumps were ejected at a different angle so that they have time to reach elevations of $\sim 1$ au at the observed projected distances. Note the correlation between the blowout angle and $\beta$ (argument of pericenter and blowout force/gravitational force in Figure \ref{fig:mcmc}).

\section{Summary and Conclusions}
\label{sec:summary}

We present a new mechanism to produce coherent dust structures accelerating radially outward, as seen in scattered light observations of the AU Mic debris disk. We appeal to gravitational interactions between an inclined, eccentric planet and an exterior debris disk with stellar wind forces to explain the dust clumps. The inclined planet kicks groups of particles strongly at apocenter and pericenter compared to elsewhere in the orbit, and eventually, when the particles are high enough, the stellar wind pushes them outward. In our code, the stellar wind ''turns on" for particles that surpass the threshold $z_h$, meaning particles within the midplane will not be pushed out by the stellar wind. This detail allowed us to avoid having to continuously replenish the disk at every timestep.  This model naturally provides a mechanism for creating periodic dust clump ejections.

We find that to produce a periodic pattern of particle ejection that aligns to the planet's orbital period, the disk's radial inner edge should be less than approximately twice the semi-major axis of the planet. We find that more than one planet-disk configuration is capable of emitting the majority of clumps on one side of the disk, though fitting for their observed measured velocity reduces the parameter space. With an MCMC and Nbody simulation, we found that a planet at $\approx$ 3.5 au, disk inner edge of $\approx$ 5.4 au, and $\beta \approx 1.8$ produces puffs that reach heights and velocities comparable to those observed in scattered light. With surface density maps, we find that this model produces a morphology that matches observations; specifically, the clumps become dimmer and elongated over time, and we see arch-like features.
We anticipate that future planet hunting campaigns will be able to further validate or disprove the existence of a planet at this distance.

The model discussed in this paper can be considered in the context of the broader population of debris disks. As high resolution observations of debris disks in both scattered light and thermal emission increase, this model may be helpful in explaining other asymmetric structures, including the ``Cat's tail'' in $\beta$ Pic \citep{Rebollido2024}. 

\begin{acknowledgments}
We thank the anonymous reviewer for comments that helped improve the discussion in this paper. All authors acknowledge support from NASA's Interdisciplinary Consortia for Astrobiology Research (NNH19ZDA001N-ICAR) under grant number 80NSSC21K0597. A.H.R. thanks the LSSTC Data Science Fellowship Program, which is funded by LSSTC, NSF Cybertraining Grant \#1829740, the Brinson Foundation, and the Moore Foundation; her participation in the program has benefited this work. This material is based upon work supported by the National Science Foundation Graduate Research Fellowship Program under Grant No. 1842400. Any opinions, findings, and conclusions or recommendations expressed in this material are those of the author(s) and do not necessarily reflect the views of the National Science Foundation. A.H.R. and R.M.C. acknowledge support from NASA grant numbers 80NSSC21K0376 and 80NSSC23K0680. We acknowledge the use of the lux supercomputer at UC Santa Cruz, funded by NSF MRI grant AST 1828315. 
\end{acknowledgments}

\appendix
\section{Visualization at various viewing angles}
\label{sec:appendix}
\begin{figure*}
    \includegraphics[height=8.in]{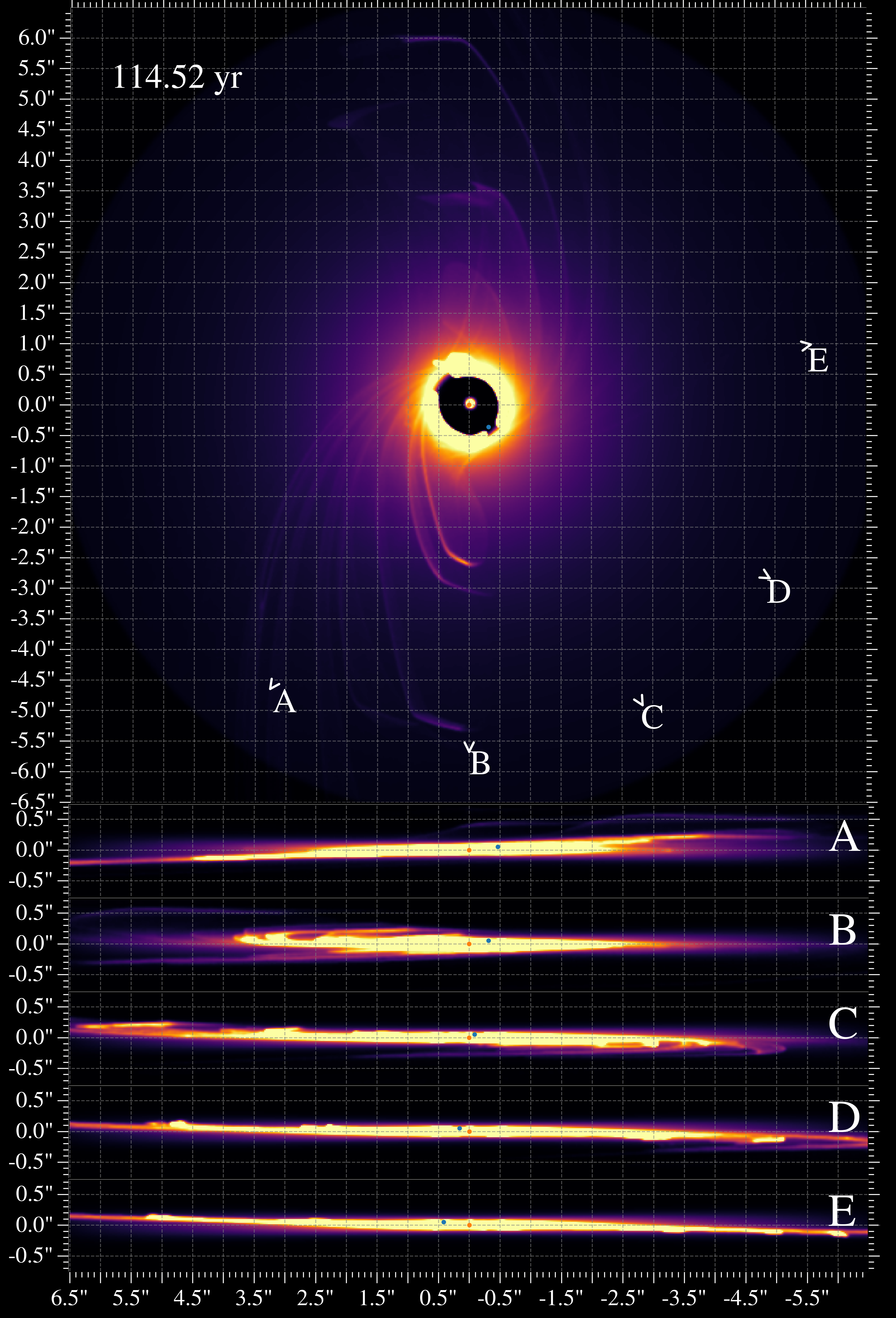}
    \caption{Animation of our simulation snapshots over 200 years, illustrating the evolution of the disk and clumps under the influence of the planet and stellar wind. The labels A-E in the top-down view (top panel) mark five observer lines of sight indicated by angled carets representing the observer eye symbol. The bottom five panels show the corresponding edge-on configurations of the disk as seen from each of those lines of sight at a given time. The apparent morphologies of the disk and clumps vary substantially depending on the viewing geometry.}
    \label{fig:disk_diff_angles}
\end{figure*}

The morphology of the disk depends on the observer's line of sight. The cross section the observer views will vary depending on the viewing angle, and the vertical separation of the particles from the disk will also vary significantly. In Figure \ref{fig:disk_diff_angles}, we show five different viewing angles for the same snapshot, denoted by letters A, B, C, D, and E. The marker, $>$, represents the observer's location at infinity with a line of sight toward the edge-on disk.
Noting the difference in geometry is interesting for comparison to future observations of debris disks that may experience a similar dynamical process.

\vspace{5mm}

\software{REBOUND: \citet{Rein2012}, REBOUNDx: \citep{Tamayo2020}, NUMPY: \citet{numpy}, MATPLOTLIB: \citet{matplotlib}, PANDAS: \citet{pandas},
            ASTROPY: \citet{astropy2013,astropy2018,astropy2022} , SCIPY: \citet{scipy} 
          }

\bibliography{aumic}
\bibliographystyle{aasjournalv7.1}

\end{document}